\documentclass[preprint,5p,twocolumn,10pt]{elsarticle}

\usepackage[pdftex,hypertexnames=false,linktocpage=true]{hyperref}
\hypersetup{pdfauthor={Lin Bolang},colorlinks=true,linkcolor=red,anchorcolor=darkgreen,citecolor=green!50!blue,filecolor=darkgreen,urlcolor=green,bookmarksnumbered=true,pdfview=FitB}
\usepackage{lipsum}
\usepackage{mathtools,cuted}
\usepackage{cuted}

\usepackage{graphicx}
\usepackage{float}
\usepackage{subfigure}

\usepackage{array}
\usepackage{booktabs} 
\usepackage{arydshln}
\usepackage{multirow}
\usepackage{makecell} 

\newcommand{\be}{\begin{equation}}  
\newcommand{\ee}{\end{equation}}  

\newcommand{\ba}{\begin{align}}  
\newcommand{\ea}{\end{align}}

\newcommand{\ra}{\rangle}
\newcommand{\la}{\langle}
\newcommand{\ma}{\Big{|}}
\newcommand{\nn}{\nonumber}

\hypersetup{colorlinks=true, citecolor=blue, urlcolor=blue, linkcolor=blue}

\usepackage{braket}

\usepackage{comment}
\begin{document}

\begin{frontmatter}
\title{Generalized parton distributions of valence, sea, and gluon components of the proton}

\author[imp,ucas]{Yiping~Liu}
\ead{liuyiping@impcas.ac.cn}

\author[imp,ucas,iowa]{Siqi~Xu}
\ead{xsq234@impcas.ac.cn}

\author[imp,ucas]{Chandan~Mondal\corref{cor1}}
\ead{mondal@impcas.ac.cn}

\author[imp,ucas]{Xingbo~Zhao}
\ead{xbzhao@impcas.ac.cn}

\author[iowa]{James~P.~Vary}
\ead{jvary@iastate.edu}

\author[]{\\\vspace{0.2cm}(BLFQ Collaboration)}

\address[imp]{Institute of Modern Physics, Chinese Academy of Sciences, Lanzhou, Gansu, 730000, China}
\address[ucas]{School of Nuclear Physics, University of Chinese Academy of Sciences, Beijing, 100049, China}

\address[iowa]{Department of Physics and Astronomy, Iowa State University, Ames, IA 50011, USA}
\cortext[cor1]{Corresponding author}

\begin{abstract}
We compute the generalized parton distributions (GPDs) of valence quarks, sea quarks, and gluons in the proton using light-front wave functions obtained within the basis light-front quantization (BLFQ) framework, providing a realistic description of the nucleon at a low resolution scale. The wave functions are derived from a light-front QCD Hamiltonian without an explicit confining potential and include the three-quark, three-quark–gluon, and three-quark–quark–antiquark Fock sectors. For the first time within BLFQ, we evaluate quark GPDs at nonzero skewness in both the DGLAP and ERBL regions, while gluon GPDs are computed in the DGLAP region. The resulting GPDs exhibit qualitative features similar to, but smaller than the ${\rm GUMP\,1.0}$ global extraction of GPDs based on experimental and lattice QCD data at next-to-
leading order accuracy. We further compute the associated Compton form factors and obtain results consistent with the global analysis.
\end{abstract}
\begin{keyword}
GPDs \sep Proton \sep Valence quarks \sep Gluons \sep Sea quarks \sep Light-front quantization
\end{keyword}
\end{frontmatter}

\section{Introduction}
Extensive experimental and theoretical studies over the past decades have revealed that the nucleon is far more complex than a simple system of three quarks bound by gluons. Rather, it is a dynamical many-body system composed of valence quarks, sea quarks, and gluons—collectively referred to as partons—that interact and move relative to one another according to the principles of QCD. Over the past two decades, substantial experimental and theoretical efforts have been devoted to the study of generalized parton distributions (GPDs), which provide crucial insights into the three-dimensional spatial distribution, spin, and orbital motion of partons inside the nucleon (see Refs.~\cite{Diehl:2023nmm,Lorce:2025aqp,Boer:2025ixc} and references therein). 

GPDs have attracted considerable attention because they unify and extend conventional observables while encoding far richer structural information. In the forward limit, GPDs reduce to the collinear parton distribution functions (PDFs), which describe the longitudinal momentum distributions of quarks and gluons inside the nucleon. Conversely, integrating GPDs over the longitudinal momentum fraction yields the electromagnetic form factors, which characterize the nucleon’s charge and magnetization distributions.

Precise knowledge of GPDs is essential for the analysis and interpretation of a wide range of hard exclusive scattering processes. These include deeply virtual Compton scattering (DVCS)~\cite{Ji:1996nm,Goeke:2001tz}, deeply virtual meson production (DVMP)~\cite{Goloskokov:2007nt,Collins:1996fb,Goloskokov:2024egn}, neutrino and electroweak meson production~\cite{Pire:2017yge,Pire:2021dad}, single diffractive hard exclusive processes (SDHEPs)~\cite{Qiu:2022pla,Grocholski:2022rqj,Duplancic:2022ffo}, timelike Compton scattering (TCS)~\cite{Berger:2001xd}, wide-angle Compton scattering (WACS)~\cite{Radyushkin:1998rt,Diehl:1998kh}, and double  deeply virtual Compton scattering (DDVCS)~\cite{Deja:2023ahc}.

Extensive experimental programs worldwide have made substantial progress in probing GPDs, including measurements by Hall A~\cite{JeffersonLabHallA:2006prd,JeffersonLabHallA:2007jdm} and CLAS~\cite{CLAS:2001wjj,CLAS:2006krx,CLAS:2007clm} at Jefferson Lab (JLab); ZEUS~\cite{ZEUS:1998xpo,ZEUS:2003pwh}, H1~\cite{H1:2001nez,H1:2005gdw}, and HERMES~\cite{HERMES:2001bob,HERMES:2006pre,HERMES:2008abz} at DESY; and COMPASS~\cite{dHose:2004usi} at CERN.
The precise determination of proton GPDs is a central objective of both current and future experimental facilities, including the Electron–Ion Colliders (EICs)~\cite{Accardi:2012qut,AbdulKhalek:2021gbh,Anderle:2021wcy,AbdulKhalek:2022hcn,Abir:2023fpo,Amoroso:2022eow,Hentschinski:2022xnd}, the Large Hadron–Electron Collider (LHeC)~\cite{LHeCStudyGroup:2012zhm,LHeC:2020van}, and the 12~GeV upgrade at JLab~\cite{Dudek:2012vr,Burkert:2018nvj,Accardi:2023chb}.

From a theoretical perspective, a variety of nonperturbative approaches have been employed to study valence quark GPDs, often within phenomenological frameworks~\cite{Pasquini:2005dk,Pasquini:2006dv,Meissner:2009ww,Boffi:2002yy,Scopetta:2003et,Choi:2001fc,Choi:2002ic,Kaur:2023lun}. Among the most promising first-principles methods for determining GPDs is Euclidean lattice QCD~\cite{Ji:2013dva,Ji:2020ect,Lin:2021brq,Lin:2020rxa,Bhattacharya:2022aob,Alexandrou:2021bbo,Alexandrou:2022dtc,Guo:2022upw,Alexandrou:2020zbe,Gockeler:2005cj,QCDSF:2006tkx,Alexandrou:2019ali,Hannaford-Gunn:2024aix}.

Although gluon and sea-quark GPDs have been explored less extensively than their valence-quark counterparts, recent studies have increasingly focused on both chiral-even and chiral-odd gluon GPDs. These investigations employ a range of modern techniques, including light-front spectator models~\cite{Tan:2023kbl,Chakrabarti:2024hwx,Chakrabarti:2023djs}, basis light-front quantization (BLFQ)~\cite{Lin:2023ezw,Lin:2024ijo,Zhang:2025nll}, light-front holography~\cite{Gurjar:2022jkx,deTeramond:2021lxc}, double-distribution representations~\cite{Goloskokov:2024egn}, and holographic string-based approaches~\cite{Mamo:2024jwp,Mamo:2024vjh} to study these leading-twist phenomena. Meanwhile, sea-quark GPDs have been investigated within light-cone model frameworks~\cite{Choudhary:2023unw,Luan:2023lmt}.

In this work, we determine the mass eigenstates of the light-front QCD (LFQCD) Hamiltonian within the BLFQ framework~\cite{Vary:2009gt,Vary:2025yqo}. The Hamiltonian explicitly includes quark (\(q\)) and gluon (\(g\)) degrees of freedom and incorporates the essential QCD interactions relevant for the dominant nucleon Fock sectors: \( |qqq\rangle \), \( |qqqg\rangle \), and \( |qqqq\bar{q}\rangle \)~\cite{Brodsky:1997de}. The resulting light-front wave functions (LFWFs), obtained as the Hamiltonian eigenvectors, are then employed to compute the GPDs of valence quarks, gluons, and sea quarks in the proton. By extending the approach to include the \( |qqqq\bar{q}\rangle \) Fock component, we enable the evaluation of quark GPDs at nonzero skewness in both the  Dokshitzer-Gribov-Lipatov-Altarelli-Parisi (DGLAP) and Efremov-Radyushkin-Brodsky-Lepage (ERBL)  regions, as well as a systematic study of sea-quark GPDs.

GPDs can be directly connected to experimental observables such as cross sections and beam or target spin asymmetries. DVCS is widely regarded as one of the golden channels for accessing GPDs. The amplitudes of this process are parametrized in terms of Compton form factors (CFFs), which are complex functions expressed as integrals over the longitudinal momentum fraction of the corresponding GPDs. Our approach, based on a truncated Fock space, is well suited for comparisons with experimental data at low energy scales. We subsequently apply QCD evolution to evolve the GPDs to higher momentum scales and compute the corresponding CFFs for comparison with the extracted data from a global fit of DVCS observables 
that fit neural network techniques~\cite{Moutarde:2019tqa}.



\section{LFQCD Hamiltonian and nucleon LFWFs}
The internal structure of the nucleon is encoded in its LFWFs, which are obtained by solving the light-front Hamiltonian eigenvalue equation
\begin{align}
P^- P^+ |\Psi\rangle = M^2 |\Psi\rangle ,
\end{align}
where \(P^+ = P^0 + P^3\) is the longitudinal momentum, \(P^- = P^0 - P^3\) is the light-front Hamiltonian, and \(M^2\) is the invariant mass squared of the nucleon. At fixed light-front time, the nucleon state admits a Fock-space expansion,
\begin{align}\label{eq:Fock_space}
|\Psi\rangle =\,& \psi^{(3q)}|qqq\rangle + \psi^{(3q+g)}|qqqg\rangle \nonumber\\
&+ \sum_{q=u,d,s} \psi^{(3q+q\bar{q})}|qqqq\bar{q}\rangle + \dots ,
\end{align}
where the coefficients \(\psi^{(\cdots)}\) represent the probability amplitudes of the corresponding partonic configurations. These amplitudes define the LFWFs, which provide the basis for computing nucleon-structure observables.

The LFQCD Hamiltonian in the light-front gauge $A^+ = 0$, including
the interactions relevant to the Fock sectors explicitly listed in Eq.~\eqref{eq:Fock_space}, is given
by~\cite{Brodsky:1997de}
\begin{align}\label{eqn:PQCD}
P^-_{\rm QCD}=&\int {\rm d}^2x^\perp {\rm d}x^-\Big\{\frac{1}{2}\bar{\Phi}\gamma^+\frac{(m_q+\delta m_q)^2+(i\partial^\perp)^2}{i\partial^+}\Phi\nonumber\\
&+\frac{1}{2}A_a^i[\delta m_{g}^2+(i\partial^\perp)^2]A_a^i+g_s\bar{\Phi}\gamma_\mu T^aA_a^\mu\Phi\nonumber\\
&+\frac{1}{2}g_s^2\bar{\Phi}\gamma^+T^a\Phi\frac{1}{(i\partial^+)^2}\bar{\Phi}\gamma^+T^a\Phi\nonumber\\
&+\frac{g_s^2C_F}{2}\bar{\Phi}\gamma_\mu A^\mu\frac{\gamma^+}{i\partial^+}\gamma_\nu A^\nu\Phi\Big\},
\end{align}
Here, $\Phi$ and $A^\mu$ denote the quark and gluon fields, respectively,
$T^a$ are the $SU(3)$ color generators, and $\gamma^\mu$ are the Dirac matrices.
The first two terms correspond to the kinetic energies of quarks and gluons,
while the remaining terms represent vertex and instantaneous interactions
governed by the strong coupling constant $g_s$.
Ultraviolet divergences are regulated using a Fock-sector–dependent renormalization
scheme~\cite{Zhao:2014hpa,Lan:2021wok,Xu:2022yxb,Xu:2024sjt}, where $\delta m_q$
and $\delta m_g$ denote the quark and gluon mass counterterms, respectively.
In addition, a separate quark mass parameter $m_f$ is introduced in the vertex
interactions to model nonperturbative effects~\cite{Burkardt:1998dd,Glazek:1992aq}.

We employ the BLFQ framework~\cite{Vary:2009gt,Vary:2025yqo}
to solve the light-front Hamiltonian eigenvalue problem. Over the past decade, BLFQ has been
progressively developed for studies of nucleon structure. Early applications focused on the
leading Fock sector $|qqq\rangle$ using an effective Hamiltonian with explicit confinement interaction and
one-gluon exchange interaction, successfully reproducing key nucleon observables~\cite{Mondal:2019jdg,Xu:2021wwj}.
Subsequent extensions incorporated the $|qqqg\rangle$ sector with QCD-motivated interactions
replacing the one-gluon exchange interaction, enabling detailed investigations of gluon contributions
to nucleon structure, including helicity, orbital angular momentum, gravitational form
factors, and three-dimensional imaging via GPDs and TMDs~\cite{Xu:2022yxb,Nair:2025sfr,Lin:2024ijo,
Lin:2023ezw,Zhang:2025nll,Yu:2024mxo,Zhu:2024awq}.

Most recently, BLFQ has achieved fully relativistic, nonperturbative solutions of the QCD
Hamiltonian without an explicit confining potential while including the $|qqq\rangle$,
$|qqqg\rangle$, and $|qqqq\bar{q}\rangle$ sectors. These calculations yield predictions for
quark and gluon matter densities, helicity and transversity distributions, and spin-related
observables that show encouraging agreement with experimental and phenomenological
results~\cite{Xu:2024sjt}. Together, these advances represent significant
progress toward a first-principles understanding of nucleon structure within the BLFQ
framework.

For the BLFQ basis, we adopt plane-wave states in the longitudinal direction defined within
a one-dimensional box of length $2L$, with antiperiodic (periodic) boundary
conditions for quarks (gluons). In the transverse plane, we employ two-dimensional
harmonic oscillator (2D-HO) wave functions $\Phi_{nm}(\vec{p}_\perp; b)$ with scale
parameter $b$, together with light-cone helicity states in spin space~\cite{Zhao:2014xaa}.
This choice of basis transforms the light-front Hamiltonian eigenvalue problem into a
matrix eigenvalue problem.

Each single-parton basis state in momentum and spin space is labeled by the quantum
numbers $\bar{\alpha} = \{k, n, m, \lambda\}$. The longitudinal quantum number $k$
determines the light-front momentum $p^+ = 2\pi k/L$, taking positive half-integer
(integer) values for quarks (gluons); the gluon zero mode is omitted. The transverse
structure is specified by the radial quantum number $n$ and orbital angular momentum $m$.
The light-cone helicity $\lambda$ is used to specify the spin degree of freedom. For Fock sectors containing multiple color-singlet
configurations, additional labels are introduced to distinguish them; in particular,
the $|qqqg\rangle$ and $|qqqq\bar{q}\rangle$ sectors contain two and three independent
color-singlet states, respectively.

Two truncation parameters, $N_{\rm max}$ and $K$, are introduced to render the BLFQ
basis finite~\cite{Zhao:2014xaa}. The parameter $N_{\rm max}$ limits the total transverse
excitation of the 2D-HO basis according to
$\sum_i (2n_i + |m_i| + 1) \leq N_{\rm max}$. The parameter $K$ fixes the total
longitudinal momentum, $\sum_i k_i = K$, where the longitudinal momentum fraction of
each parton is given by $x_i = k_i / K$. Consequently, $K$ controls the longitudinal
resolution and, hence, the resolution of PDFs.

The resulting nucleon LFWFs with helicity $\Lambda$ are
expressed, within each Fock sector, as
\begin{equation}
\Psi^{\mathcal{N},\,\Lambda}_{\{x_i,\vec{p}_{\perp i},\lambda_i\}}
=
\sum_{\{n_i m_i\}}
\psi^{\mathcal{N}}(\{\bar{\alpha}_i\})
\prod_{i=1}^{\mathcal{N}}
\phi_{n_i m_i}(\vec{p}_{\perp i},b),
\label{eqn:wf}
\end{equation}
where $\mathcal{N}$ denotes the number of constituents in the Fock sectors
$|qqq\rangle$, $|qqqg\rangle$, and $|qqqq\bar{q}\rangle$, and
$\psi^{\mathcal{N}}(\{\bar{\alpha}_i\})$ are the corresponding BLFQ amplitudes.

\section{Generalized parton distributions\label{Sec3}}
The unpolarized quark and gluon GPDs are defined through off-forward matrix elements of the corresponding bilocal operators between proton states~\cite{Diehl:2003ny}: 
\begin{align}
F^{q}_{\Lambda,\Lambda'} &= \int \frac{dz^-}{4\pi} e^{ix\bar{P}^+z^-} \nn \\
 &\times \la P',\Lambda' \ma \bar{\Phi}\left(-\frac{z}{2}\right)\,\gamma^+\,\Phi\left(-\frac{z}{2}\right)  \ma P,\Lambda \ra \Big{|}_{\substack{z^+=0\\z^{\perp}=0}}  \label{quark_unpol}\\
 &=  \frac{1}{2\bar{P}^+} \bar{u}(P',\Lambda')\left[ H^q \gamma^+ +  E^q  \frac{i\sigma^{+\alpha} \Delta_{\alpha}}{2M}\right] u(P,\Lambda),  \nn
 \end{align}
\begin{align}
F^{g}_{\Lambda,\Lambda'} &= \frac{1}{\bar{P}^+}\int \frac{dz^-}{2\pi} e^{ix\bar{P}^+z^-} \nn \\
 &\times \la P',\Lambda' \ma F^{+i} \left(-\frac{z}{2} \right) F^{+i}  \left(\frac{z}{2} \right)  \ma P,\Lambda \ra \Big{|}_{\substack{z^+=0\\z^{\perp}=0}}  \label{unpol}\\
&=  \frac{1}{2\bar{P}^+} \bar{u}(P',\Lambda')\left[ H^g \gamma^+ +  E^g  \frac{i\sigma^{+\alpha} \Delta_{\alpha}}{2M}\right] u(P,\Lambda),  \nn
\end{align}
where $\Phi$ denotes the quark field and $F^{+\mu}(x) = \partial^{+}A^{\mu}(x)$ corresponds to the gluon field tensor in the light-cone gauge, $A^+=0$. In the symmetric frame, the average proton momentum is defined as
$\bar{P} = \tfrac{1}{2}(P' + P)$, while the momentum transfer is $\Delta = P' - P$. Accordingly, the initial and final proton
four-momenta are given by
\begin{equation}
\begin{aligned}
P &\equiv \left((1+\xi)\bar{P}^+,\frac{M^2+\Delta_\perp^2/4}{(1+\xi)P^+},-\vec{\Delta}_\perp/2\right),\\
P^{\prime} &\equiv \left((1-\xi)\bar{P}^+,\frac{M^2+\Delta_\perp^2/4}{(1-\xi)P^+},\vec{\Delta}_\perp/2\right). \label{Ppp}
\end{aligned}
\end{equation}
Note that the GPDs depend on the variables $x$, $\xi=- \Delta^+/2\bar{P}^+$, and $t= \Delta^2$. From the expression for $\Delta^-$, one can explicitly derive the following relation
\begin{align}
- t= \frac{4 \xi^2 M^2 + \Delta_\perp^2}{(1-\xi^2)}\,. \label{mt_def}
\end{align}

In a reference frame where the proton momenta $\vec{P}$ and
$\vec{P}^{\,\prime}$ lie in the $x$--$z$ plane~\cite{Pasquini:2005dk},
we explicitly derive the relations for the GPDs as follows:
\begin{equation}
    \begin{aligned}
        H^{q/g}(x, \xi, t)=&\frac{1}{\sqrt{1-\xi^2}} F^{q/g}_{++}+\frac{2 M \xi^2}{\sqrt{1-\xi^2} \Delta_{\perp 1}} F^{q/g}_{-+}, \\
        E^{q/g}(x, \xi, t)=&\frac{2 M \sqrt{1-\xi^2}}{\Delta_{\perp 1}}  F^{q/g}_{-+}, 
    \end{aligned}
\end{equation}
where the proton helicity  is denoted by $\Lambda=+(-)$, corresponding to $+1(-1)$, respectively.

In the overlap representation, only diagonal overlaps between identical
$\mathcal{N}$-particle Fock states contribute in the DGLAP region,
$\xi < |x| < 1$. Meanwhile, in the ERBL region, $0 < |x| < \xi$, the GPDs receive contributions from
non-diagonal overlaps between different Fock sectors, specifically
transitions of the form $\mathcal{N} \to \mathcal{N}+2$.
In these different kinematic regimes, the correlator
$F^{q/g}_{\Lambda\Lambda'}$ can be expressed in terms of the LFWFs as follows:
\begin{align}
F^{q/g}_{\Lambda^\prime,\Lambda}=&\sum_{\{\mathcal{N}, \lambda_i\}} \int \left[{\rm d}\mathcal{X} \,{\rm d}\mathcal{P}_\perp\right]_{\rm DGLAP}\, \Psi^{\mathcal{N},\Lambda^\prime *}_{\{x^{\prime\prime}_i,\vec{k}^{\prime\prime}_{i\perp},\lambda_i\}}\Psi^{\mathcal{N},\Lambda}_{\{x_i^{\prime},\vec{k}_{i\perp}^{\prime},\lambda_i\}},\\
F^{q/g}_{\Lambda^\prime,\Lambda}=&\sum_{\{ \lambda_i\}} \int \left[{\rm d}\mathcal{X} \,{\rm d}\mathcal{P}_\perp\right]_{\rm ERBL}\, \Psi^{\mathcal{N},\Lambda^\prime *}_{\{x^{\prime\prime}_i,\vec{k}^{\prime\prime}_{i\perp},\lambda_i\}}\Psi^{\mathcal{N}+2,\Lambda}_{\{x_i^{\prime},\vec{k}_{i\perp}^{\prime},\lambda_i\}}\nonumber \\&\times\delta_{\lambda_{\mathcal{N}+1},-\lambda_{\mathcal{N}+2}},
\end{align}
where we use the abbreviation for the DGLAP region,
\begin{align}
&\left[{\rm d}\mathcal{X} \,{\rm d}\mathcal{P}_\perp\right]_{\rm DGLAP}=(\sqrt{x^2-\xi^2})^l(\sqrt{1-\xi^2}\ )^{2-n}\nonumber\\
\times &\prod_{i=1}^{\mathcal{N}}\left[\frac{{\rm d}x_i{\rm d}^2 \vec{k}_{i\perp}} {16\pi^3}\right]\delta(x-x_1)\, \nonumber\\
\times &16 \pi^3 \delta \left(1-\sum_{i=1}^{\mathcal{N}} x_i\right) \delta^2 \left(\sum_{i=1}^{\mathcal{N}}\vec{k}_{i\perp}\right), 
\end{align}
and for the ERBL region,
\begin{align}
&\left[{\rm d}\mathcal{X} \,{\rm d}\mathcal{P}_\perp\right]_{\rm ERBL}=(\sqrt{\xi^2-x^2})^l(\sqrt{1-\xi^2}\ )^{2-n}\nonumber\\
\times &\prod_{i=1}^{\mathcal{N}+2} \left[\frac{{\rm d}x_i{\rm d}^2 \vec{k}_{i\perp}} {16\pi^3}\right]\delta(x-x_{\mathcal{N}+1})\, \nonumber\\
&\times16 \pi^3 \delta \left(1-\xi-\sum_{i=1}^{\mathcal{N}} x_i\right) \delta^2 \left(\sum_{i=1}^{\mathcal{N}+2}\vec{k}_{i\perp}-\frac{\vec{\Delta}_\perp}{2}\right)\,\\
&\times \delta \left(2\xi- x_{\mathcal{N}+1}-x_{\mathcal{N}+2}\right)\, \delta^2 \left(\vec{k}_{\mathcal{N}+1\perp}+\vec{k}_{\mathcal{N}+2\perp}-\vec{\Delta}_\perp\right),  \nonumber
\end{align} 
with $l=0\,(1)$ for quark (gluon). 
In the DGLAP region light-front momenta are $x^{\prime}_1=\frac{x_1+\xi}{1+\xi}$; $\vec{k}^{\prime}_{1\perp}=\vec{k}_{1\perp}+(1-x^{\prime})\frac{\vec{\Delta}_{\perp}}{2}$ for the initial  parton ($i=1$) and $x^{\prime}_i=\frac{x_i}{1+\xi}; ~\vec{k}^{\prime}_{i\perp}=\vec{k}_{i\perp}-{x_i^{\prime}} \frac{\vec{\Delta}_{\perp}}{2}$ for the initial spectators  ($i\ne1$) and
$x^{\prime\prime}_1=\frac{x_1-\xi}{1-\xi}$; $\vec{k}^{\prime}_{1\perp}=\vec{k}_{1\perp}-(1-x^{\prime\prime})\frac{\vec{\Delta}_{\perp}}{2}$ for the final parton and $x^{\prime\prime}_i=\frac{x_i}{1-\xi}; ~\vec{k}^{\prime}_{i\perp}=\vec{k}_{i\perp}+{x_i^{\prime\prime}} \frac{\vec{\Delta}_{\perp}}{2}$ for the final spectators. In the ERBL region, light-front momenta are $x^{\prime}_i=\frac{x_i}{1+\xi}; ~\vec{k}^{\prime}_{i\perp}=\vec{k}_{i\perp}-{x_i^{\prime}}\frac{\vec{\Delta}_{\perp}}{2}$ for the initial spectators $(i=1\rightarrow\mathcal{N})$ and $x^{\prime\prime}_i=\frac{x_i}{1-\xi}; ~\vec{k}^{\prime}_{i\perp}=\vec{k}_{i\perp}+{x_i^{\prime\prime}} \frac{\vec{\Delta}_{\perp}}{2}$ for the final spectators. For final struck partons, $x_{\mathcal{N}+1}^{\prime}=\frac{x_{\mathcal{N}+1}+\xi}{1+\xi}; x_{\mathcal{N}+2}^{\prime}=\frac{x_{\mathcal{N}+2}-\xi}{1+\xi}$. For quark GPDs, contributions originate from the $\mathcal{N}=3, 4,$ and $5$ Fock sectors, whereas gluon distributions receive contributions exclusively from the $\mathcal{N}=4$ sector. $\delta_{\lambda_{\mathcal{N}+1},-\lambda_{\mathcal{N}+2}}$ in Eq.(11) signals a helicity flip of the struck parton.

\section{Numerical result}
All calculations are performed with basis truncations $N_{\rm max}=7$ and $K=16.5$. The harmonic oscillator scale is fixed at $b=0.6$~GeV, and the ultraviolet cutoff for the instantaneous interaction is $b_{\rm inst}=2.75\pm0.05$~GeV. The model parameters
$\{m_u, m_d, m_f, g_s\}=\{1.0,\,0.85,\,5.65\pm0.05,\,2.95\pm0.05\}$ (in GeV, with $g_s$ dimensionless) are determined by fitting the proton mass and its electromagnetic properties~\cite{Xu:2024sjt}.
The relatively large constituent quark masses in the leading $|qqq\rangle$ and $|qqqg\rangle$ sectors effectively mimic confinement and generate substantial binding energy. 
In the $|qqqq\bar{q}\rangle$ sector, partons are expanded in the same basis but higher–Fock-sector–induced interactions are omitted due to the truncation. Consequently, current quark masses are used: $m_u=0.0022$~GeV, $m_d=0.0047$~GeV, and $m_s=0.094$~GeV.

At the model scale, the proton Fock-sector probabilities are $53.10\%$ ($|qqq\rangle$), $26.53\%$ ($|qqqg\rangle$), $8.52\%$ ($|qqqu\bar{u}\rangle$), $8.56\%$ ($|qqqd\bar{d}\rangle$), and $3.29\%$ ($|qqqs\bar{s}\rangle$)~\cite{Xu:2024sjt}.


At nonzero skewness, the calculation of GPDs requires overlaps of LFWFs with different longitudinal momenta. To accommodate this, we interpolate the longitudinal components of the LFWFs, which enables the extraction of all leading-twist GPDs. In this work, we focus on the chiral-even GPDs $H(x,\,\xi,\,t)$ and $E(x,\,\xi,\,t)$ along with their associated CFFs.

\begin{figure}[htp]
\centering
\includegraphics[scale=.33]{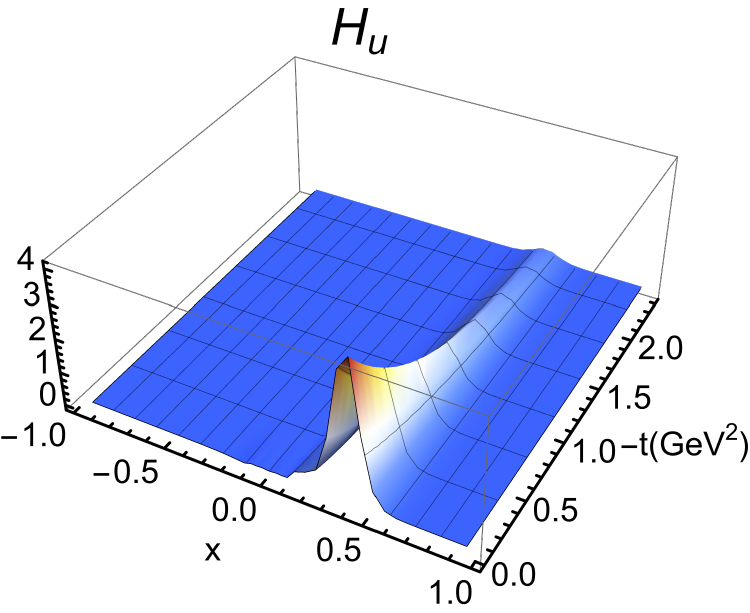}
\includegraphics[scale=.33]{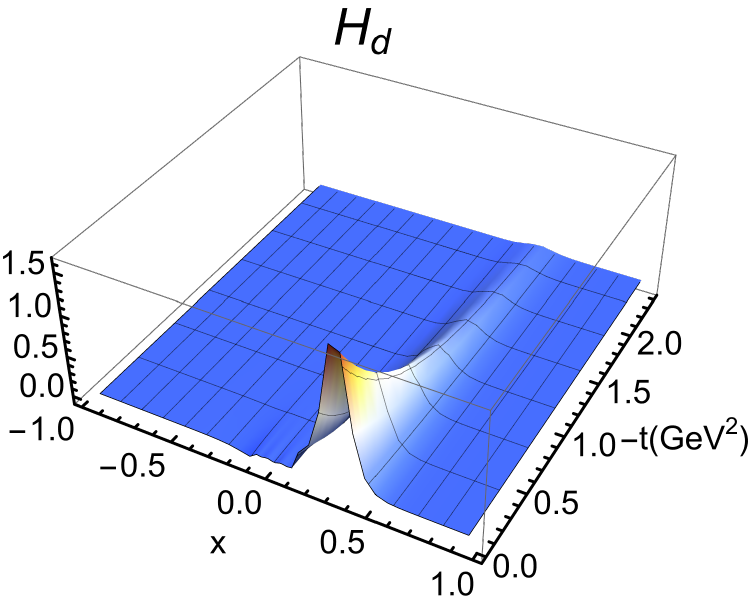}\\
\includegraphics[scale=.33]{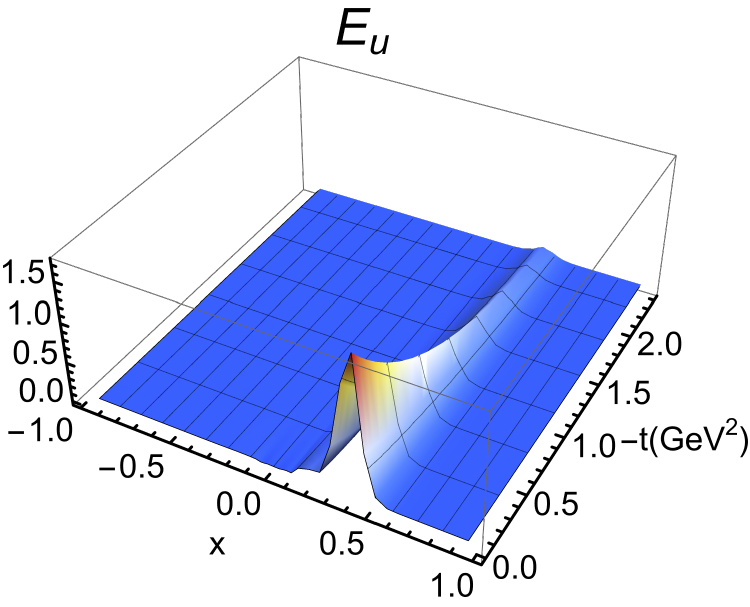}
\includegraphics[scale=.33]{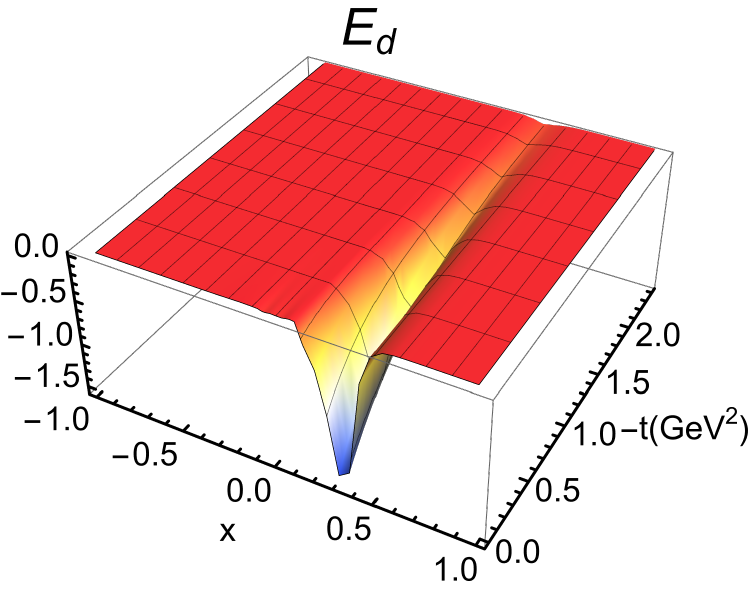}
\caption{\label{quark_full_3D} Quark GPDs \(H(x,\xi,t)\) (upper panels) and \(E(x,\xi,t)\) (lower panels) shown as functions of \(x\) and \(-t\) at fixed skewness \(\xi = 0.1\), covering both the DGLAP and ERBL regions. The left (right) panels correspond to the up (down) quark.}
\end{figure}
\begin{figure*}[htp]
\centering
\includegraphics[scale=.33]{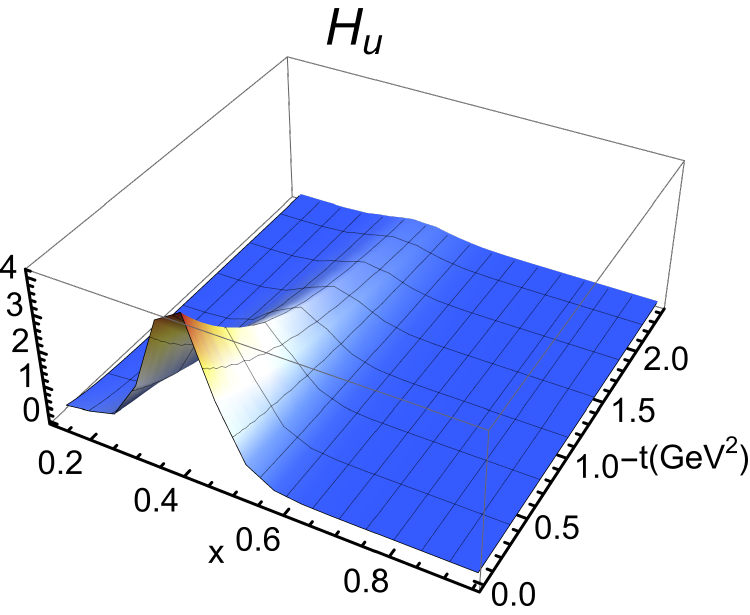}
\includegraphics[scale=.33]{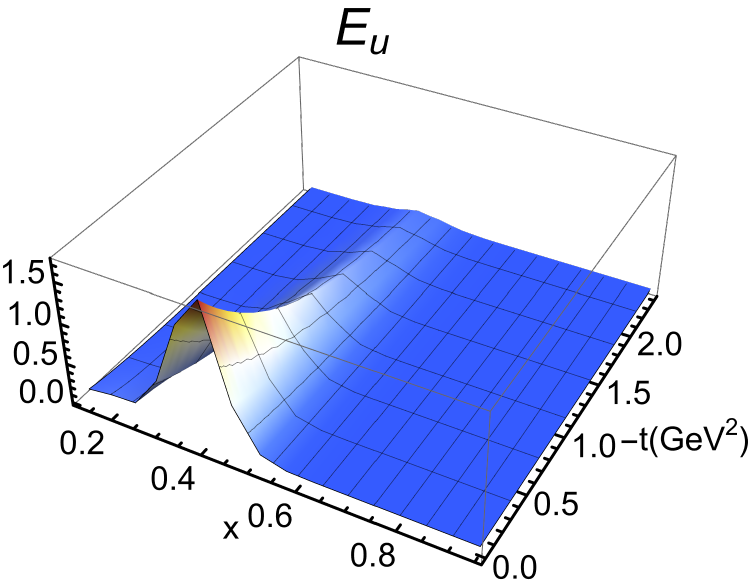}
\includegraphics[scale=.33]{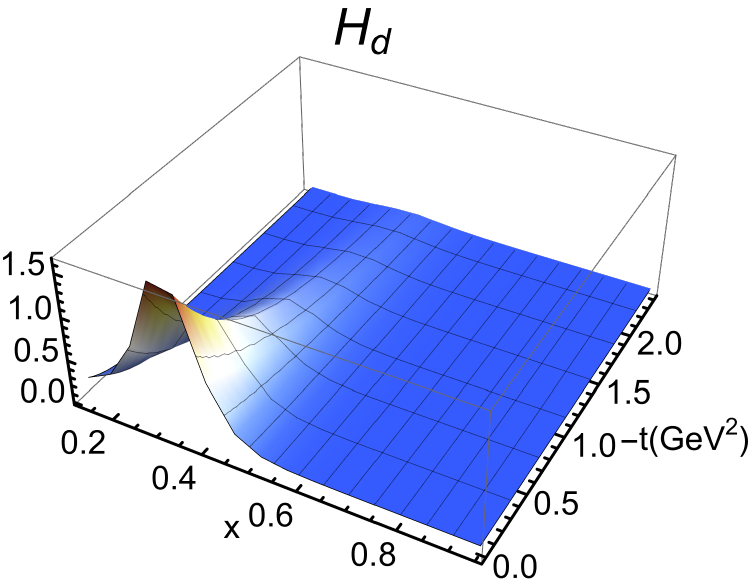}
\includegraphics[scale=.33]{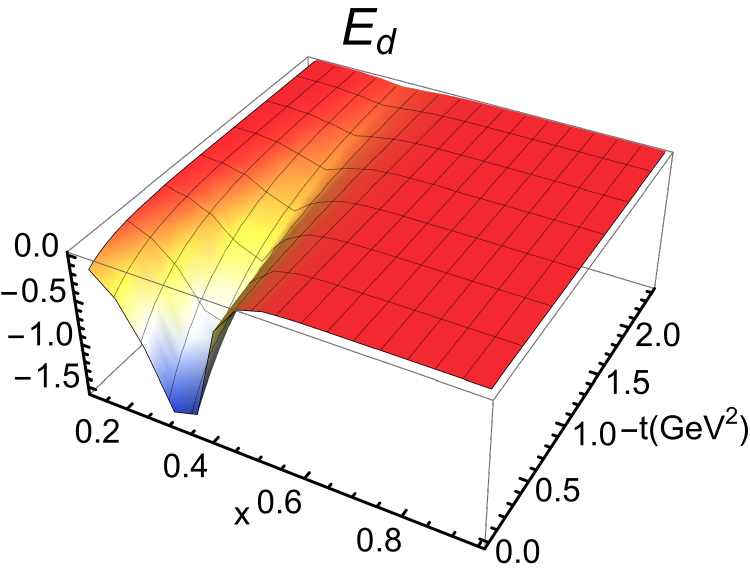}\\
\includegraphics[scale=.33]{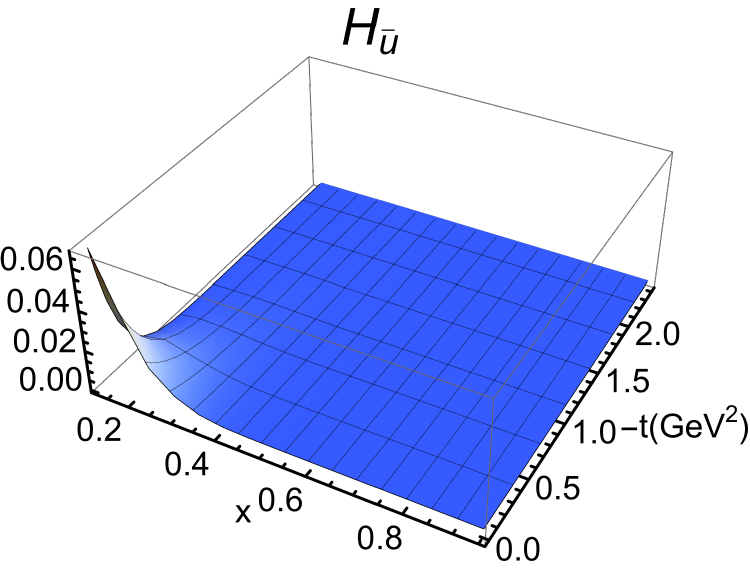}
\includegraphics[scale=.33]{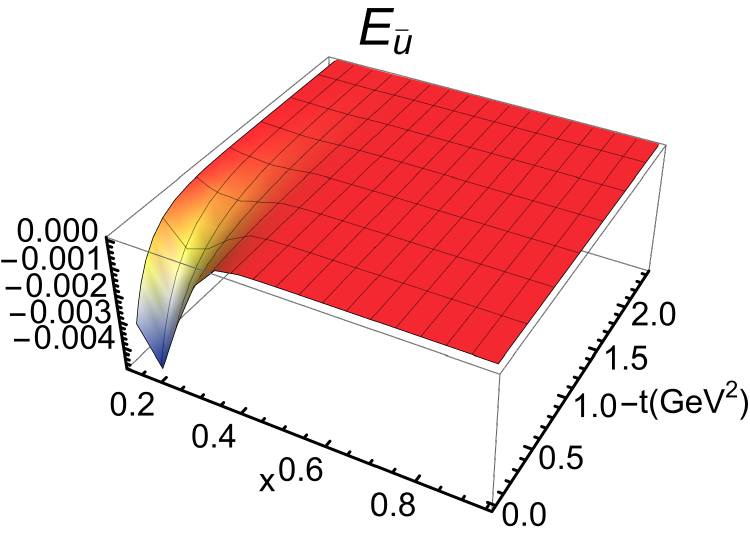}
\includegraphics[scale=.33]{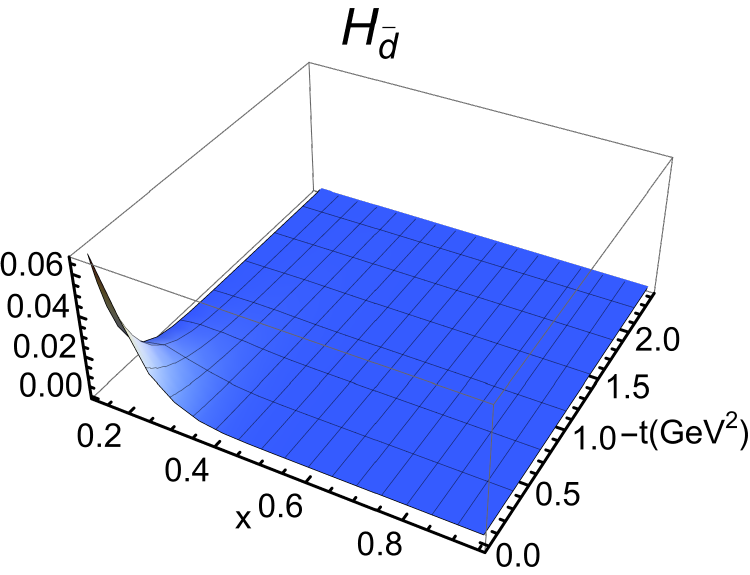}
\includegraphics[scale=.33]{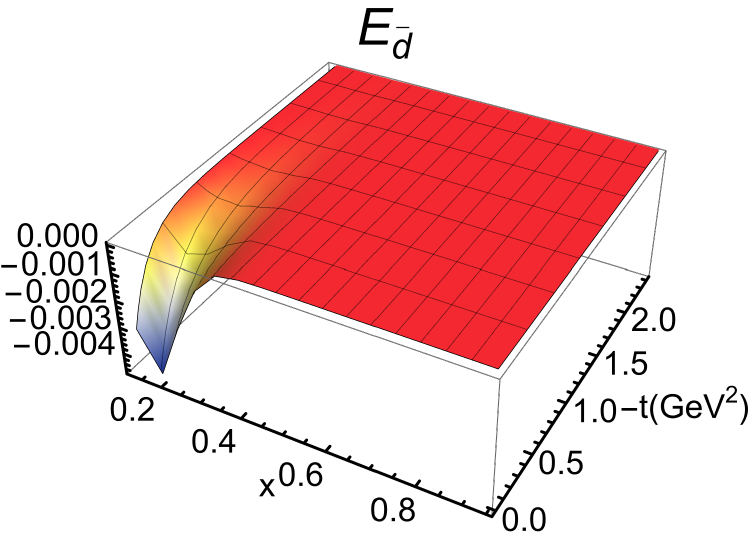}
\caption{\label{quark_DGLAP_3D} GPDs \(H(x,\xi,t)\)  and \(E(x,\xi,t)\) in DGLAP region for  quarks (upper panels) and their antiquarks (lower panels) shown as functions of \(x\) and \(-t\) at skewness $\xi=0.1$.}
\end{figure*}

\begin{figure*}[htp]
\centering
\includegraphics[scale=.33]{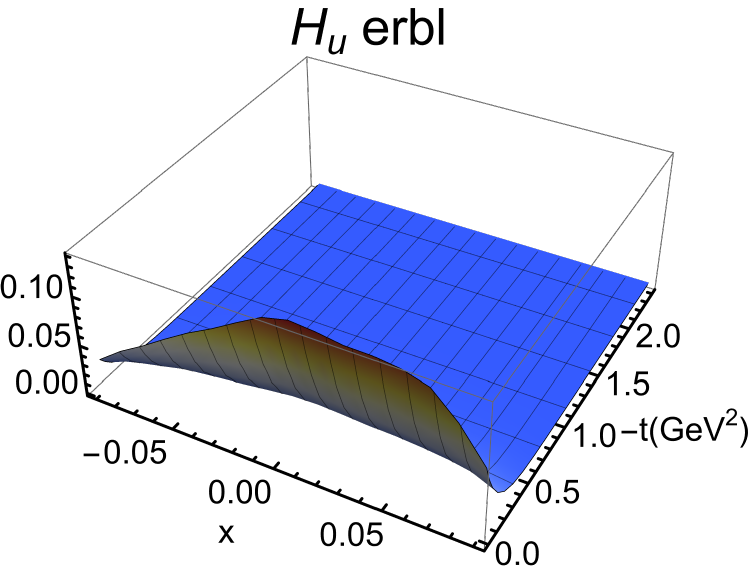}
\includegraphics[scale=.33]{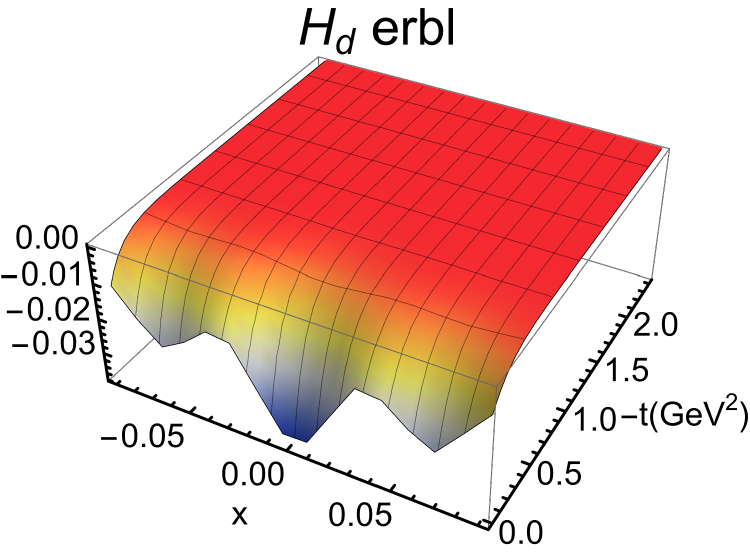}
\includegraphics[scale=.33]{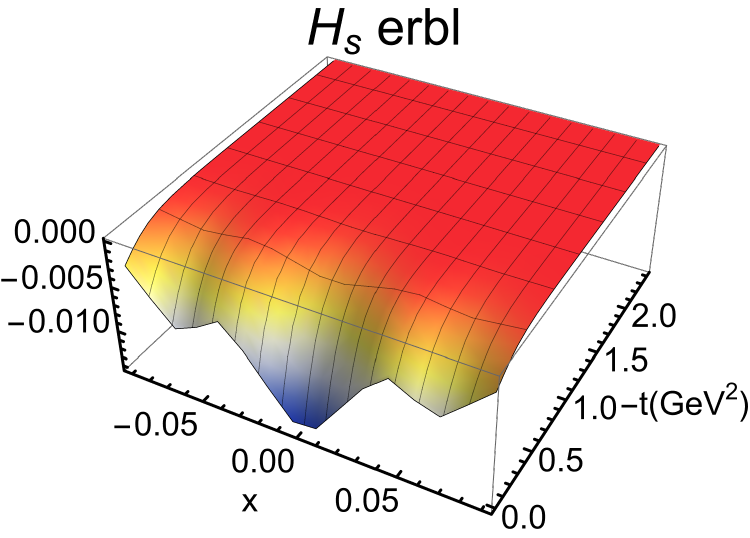}\\
\includegraphics[scale=.33]{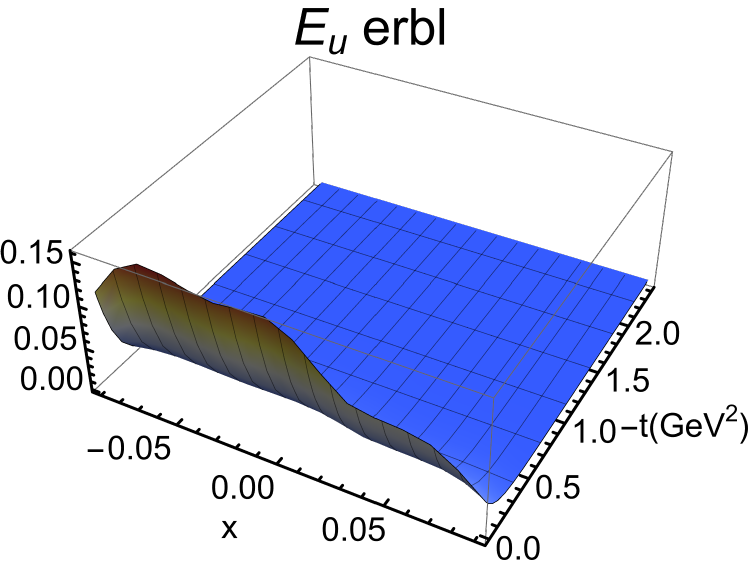}
\includegraphics[scale=.33]{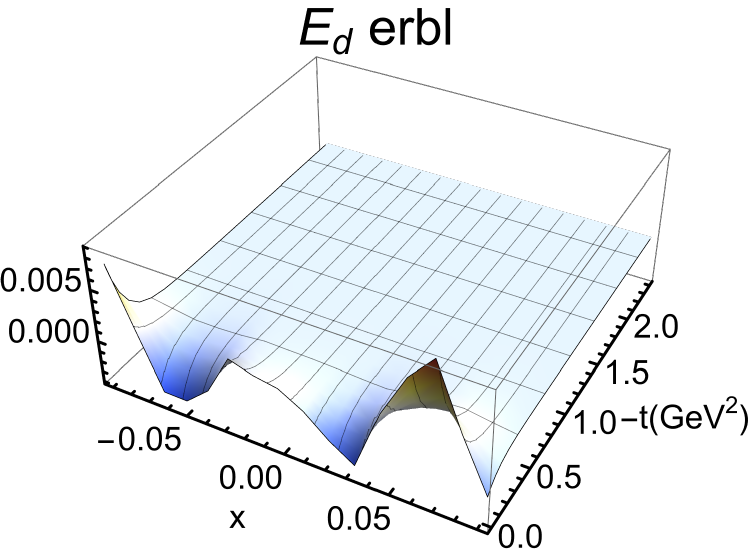}
\includegraphics[scale=.33]{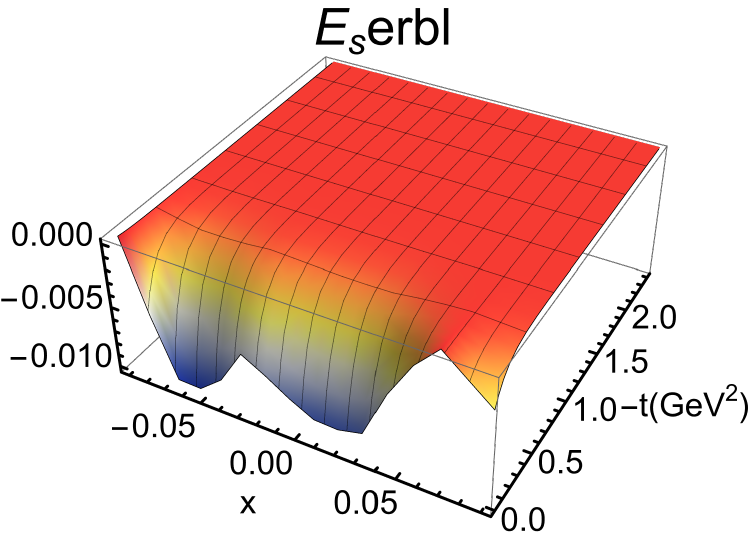}
\caption{\label{quark_ERBL_3D} Quark GPDs \(H(x,\xi,t)\) (upper panels)  and \(E(x,\xi,t)\) (lower panels) in ERBL region shown as functions of \(x\) and \(-t\) at skewness $\xi=0.1$.}
\end{figure*}

\begin{figure}[htp]
\centering
\includegraphics[scale=.33]{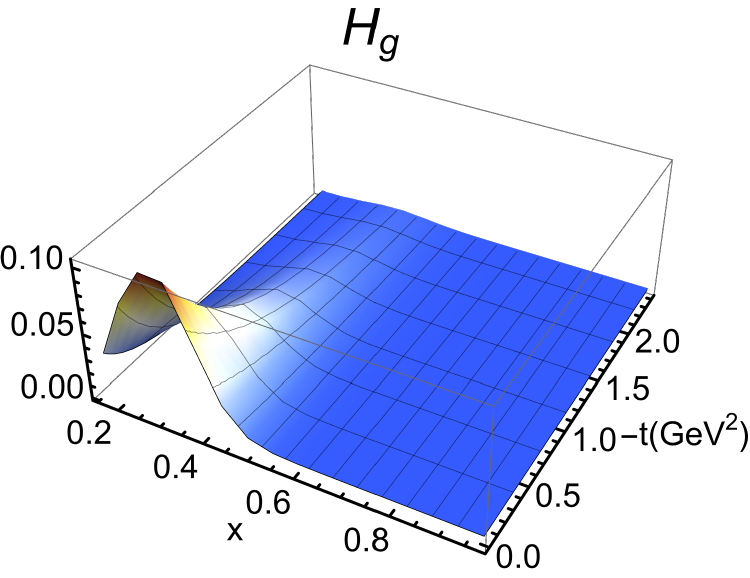}
\includegraphics[scale=.33]{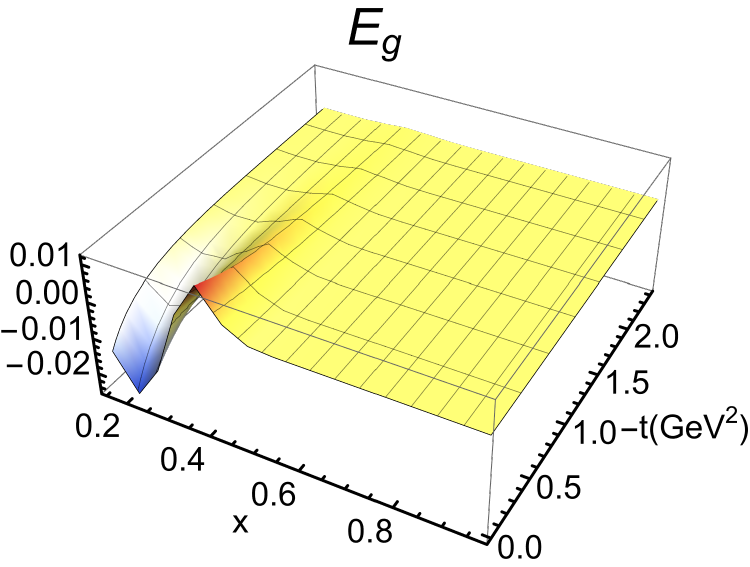}
\caption{\label{gluon_3D} Gluon GPDs \(H(x,\xi,t)\) (left panel)  and \(E(x,\xi,t)\) (right panel) shown as functions of \(x\) and \(-t\) at skewness $\xi=0.1$.}
\end{figure}

Figure~\ref{quark_full_3D} presents the quark GPDs over the full kinematic range at the model scale at skewness $\xi=0.1$. The regions are defined as follows: the quark DGLAP region corresponds to \(x>0.1\), the antiquark DGLAP region to \(x<-0.1\) (we use the definition that $F_q(-x)=-F_{\bar{q}}(x)$, where $F$ represents $H$ and $E$.), and the ERBL region to \(-0.1<x<0.1\). The dominant contribution originates from the quark DGLAP region, which is mostly contributed from $|qqq\rangle$ and $|qqqg\rangle$ Fock sectors. In contrast, both the ERBL region and the antiquark DGLAP region receive contributions from the \(|qqqq\bar{q}\rangle\) Fock sector. As this is the highest Fock sector retained in our truncated model, interactions with higher Fock sectors are omitted, leading to larger fluctuations in the corresponding wave functions compared with those of lower Fock sectors. Consequently, as \(\xi\) and \(|\vec{\Delta}_\perp|\) increase, contributions associated with the \(|qqqq\bar{q}\rangle\) sector decrease more rapidly than those from lower Fock sectors.

Figures~\ref{quark_DGLAP_3D} and~\ref{quark_ERBL_3D} provide a more detailed view of the results shown in Fig.~\ref{quark_full_3D}. Figure~\ref{quark_DGLAP_3D} displays the GPDs in the DGLAP region, with the upper panels corresponding to quark distributions and the lower panels to antiquark distributions. We observe that \(H_u\), \(H_d\), and \(E_u\) are positive, whereas \(E_d\) is negative, as expected. This sign reflects the negative anomalous magnetic moment of the down quark.

For antiquarks, both \(E_{\bar{u}}\) and \(E_{\bar{d}}\) are negative. This behavior arises because, under SU(3) symmetry, the valence contributions to \(E_u\) and \(E_d\) have opposite signs, whereas no such constraint exists for sea quarks; moreover, the \(u\bar{u}\) and \(d\bar{d}\) sea components are similar. We also note that \(H_u\) is significantly larger than the other distributions, that \(E_{\bar{q}}\) is much smaller in magnitude than \(H_{\bar{q}}\), and that \(E_{\bar{q}}\) exhibits a broader \(x\)-dependence than \(H_{\bar{q}}\). With increasing \(t\), the peaks of all GPDs shift toward larger values of \(x\).

Figure~\ref{quark_ERBL_3D} shows the GPDs in the ERBL region. Contributions in this region originate from two sources: transitions involving a valence quark and a sea antiquark, and those involving a sea quark--antiquark pair. For convenience, we refer to these as the valence and sea contributions, respectively. The first row displays the GPD \(H\). We find that \(H_d\) and \(H_s\) are nearly symmetric under \(x \to -x\), reflecting the fact that they are dominated by sea contributions, primarily in an \(s\)-wave configuration. The symmetry of the sea contribution under \(q\bar{q}\) exchange leads to this behavior. The visible fluctuations at small \(x\) further indicate the enhanced sensitivity of sea-quark wave functions in this region. In contrast, \(H_u\) has the opposite sign, since the valence contribution dominates and carries a different sign from the sea contribution.

The second row of Fig.~\ref{quark_ERBL_3D} shows the GPD \(E\). Here, \(E_d\) fluctuates in sign while \(E_u\) is positive, again reflecting the influence of the valence contribution. The sign pattern of both $H_d$ and $E_d$ differs from that observed in the DGLAP region, as the ERBL region involves different spin combinations. The distribution \(E_s\), which arises purely from sea contributions, exhibits a positive peak around \(x \simeq -0.07\) and a negative peak around \(x \simeq 0.07\), resulting from the interplay between S- and P-wave components.

Finally, Fig.~\ref{gluon_3D} shows the gluon GPDs at the initial scale. Unlike results at experimental scales, the unpolarized gluon GPD \(H_g\) does not rise at small \(x\); instead, it exhibits a peak around \(x \simeq 0.3\). This behavior indicates that the strong enhancement of gluons at small \(x\) is predominantly generated through QCD evolution. A similar trend is observed for \(E_g\), which also does not increase at small \(x\). Instead, \(E_g\) displays a small positive peak near \(x \simeq 0.4\), while its negative peak occurs at smaller \(x\), around \(x \simeq 0.2\), compared with that of \(H_g\).

\subsection{Evolved GPDs}
The GPDs obtained in our model are defined at a low hadronic scale, where nonperturbative dynamics dominates. To enable meaningful comparisons with experimental measurements and phenomenological extractions, it is therefore essential to evolve the GPDs to higher momentum scales using perturbative QCD. The scale dependence of GPDs is governed by evolution equations that generalize the DGLAP and ERBL equations~\cite{Freese:2024ypk,Bertone:2017gds,Bertone:2022frx}, with the two regimes corresponding to the regions $|x|>\xi$ and $|x|<\xi$, respectively. In the DGLAP region, the evolution describes the radiation of partons from an active quark or gluon, while in the ERBL region it accounts for the evolution of quark--antiquark pairs. This unified evolution framework preserves the polynomiality and continuity properties of GPDs across the entire kinematic domain. The evolution equations are solved using the open-source code ${\tiny{\rm APFEL^{++}}}$~\cite{Bertone:2017gds}, which is interfaced with the ${\tiny{\rm PARTONS}}$ framework. In the following, we present our results for the evolved GPDs and the corresponding CFFs at experimentally relevant scales.

\begin{figure}[htp]
\centering
\includegraphics[scale=.45]{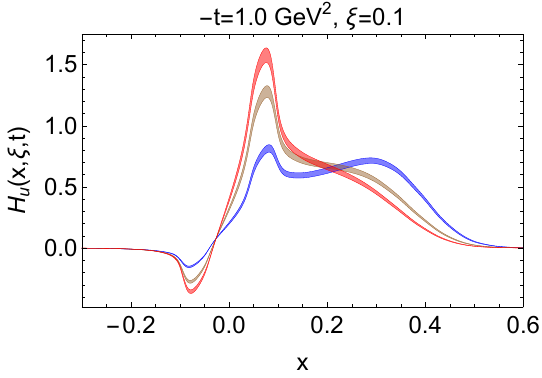}
\includegraphics[scale=.45]{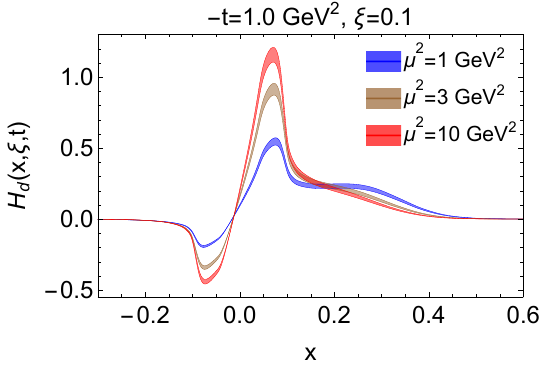}\\
\includegraphics[scale=.45]{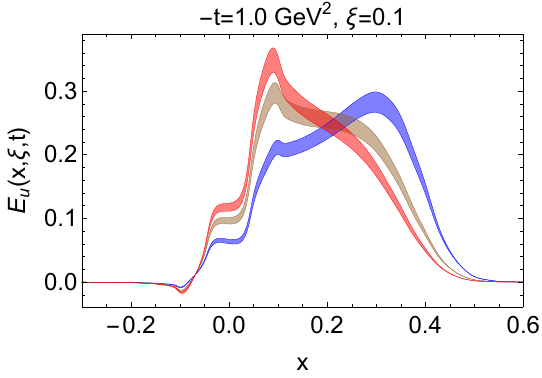}
\includegraphics[scale=.45]{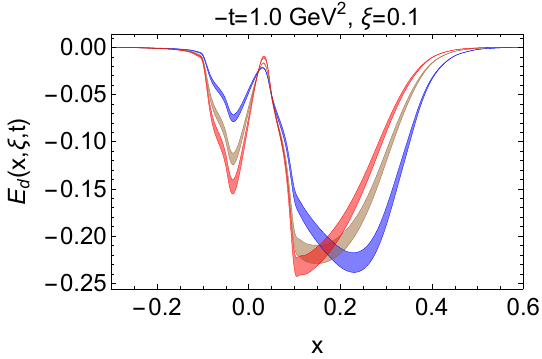}
\caption{\label{quark_full_evolved} Evolved quark GPDs \(H(x,\xi,t)\) (upper panels) and \(E(x,\xi,t)\) (lower panels) at different scale $\mu^2=1,\,3,\,10$ GeV$^2$  shown as functions of \(x\) at fixed skewness \(\xi = 0.1\) and \(-t=1\) GeV$^2$, covering both the DGLAP and ERBL regions. The left (right) panels correspond to the up (down) quark.}
\end{figure}
\begin{figure*}[htp]
\centering
\includegraphics[scale=.45]{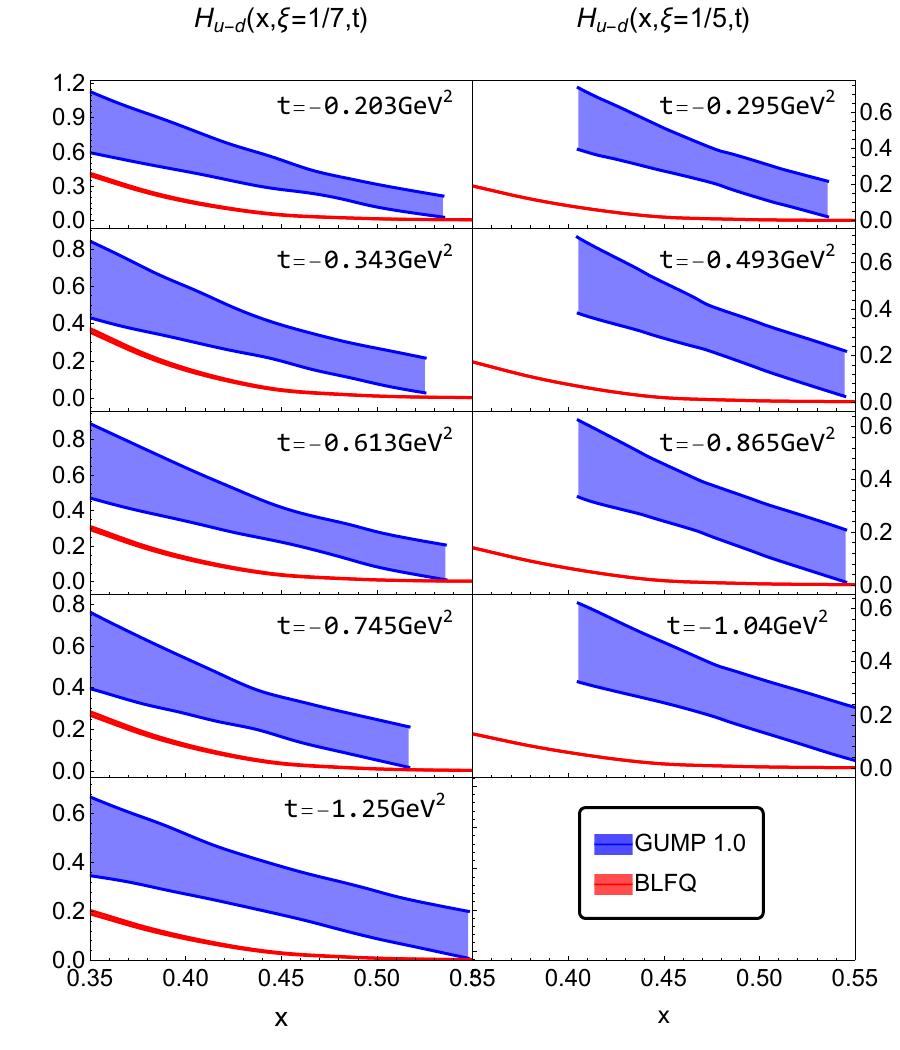}
\includegraphics[scale=.45]{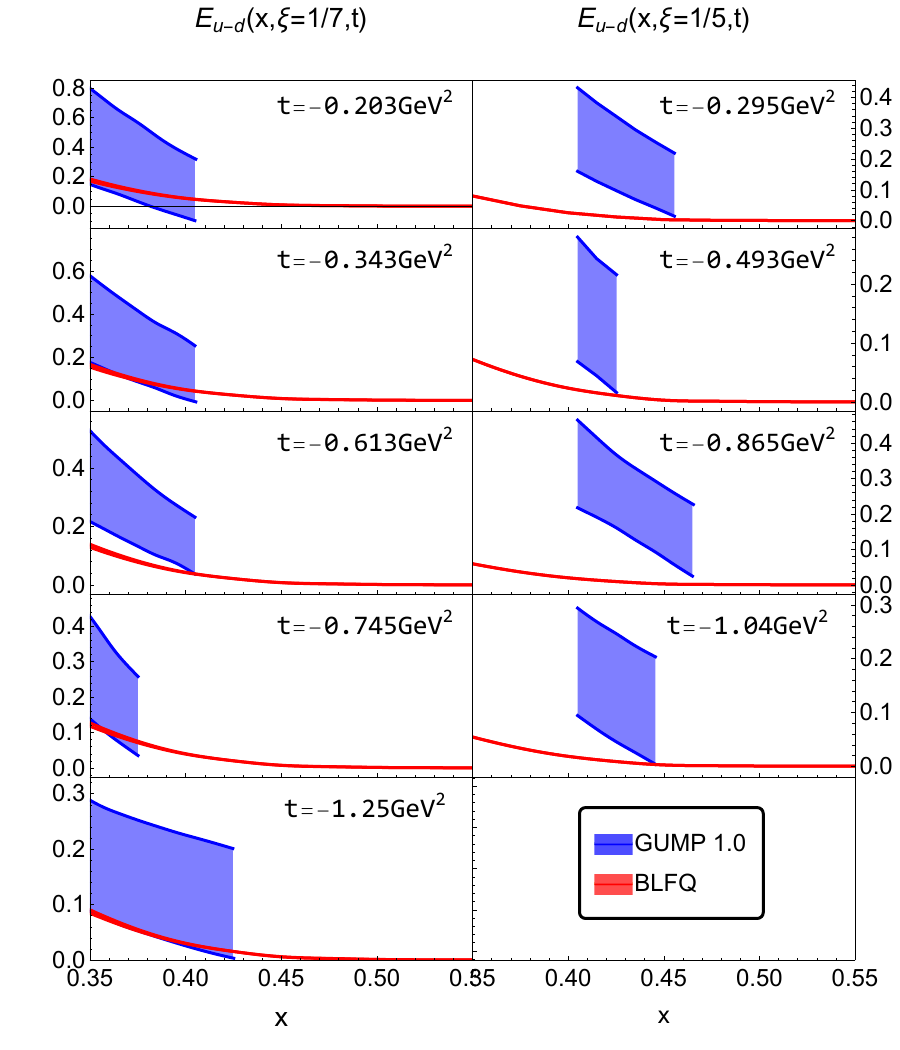}
\caption{\label{quark_gpds_compare} Comparison of the isovector combination of the quark GPDs obtained in the BLFQ framework (red bands) with the GUMP1.0 GPDs (blue bands)~\cite{Guo:2025muf} for different values of $-t$ and $\xi$ at $\mu^2=4$ GeV$^2$. The left two panels show the GPD $H_{u-d}(x,\xi,t)$ at $\xi = 1/7$ and $1/5$, respectively, while the right two panels show the GPD $E_{u-d}(x,\xi,t)$ at $\xi = 1/7$ and $1/5$, respectively.
}
\end{figure*}


Figure~\ref{quark_full_evolved} presents the quark and antiquark GPDs at fixed momentum transfer $-t=1.0~\mathrm{GeV}^2$ and skewness $\xi=0.1$, evaluated at different scales. As the scale increases, the GPDs are enhanced in the small-$x$ region and suppressed at large $x$. Consequently, the prominent peak observed at the initial scale gradually diminishes and a new peak emerges at smaller $x$, around $x\simeq 0.075$ for $H_d$, $H_u$, and $E_u$. In contrast,  $E_{d}$  does not develop a small-$x$ peak in this region due to a cancellation between valence and sea-quark contributions. We also find that the uncertainty bands estimated by varying the model parameters $g_s$, $m_f$, and $b_{\rm inst}$ are somewhat larger for the $E$ GPDs than for $H$, reflecting the stronger sensitivity of the P-wave components of the LFWFs to model parameters.

Figure~\ref{quark_gpds_compare} presents a comparison of the isovector combination of the quark GPDs obtained in the BLFQ framework with the GUMP1.0 GPDs~\cite{Guo:2025muf}. The GUMP1.0 analysis provides a consistent description of DVCS cross sections and asymmetries measured at JLab, as well as DVCS and deeply virtual $\rho$-meson production cross sections from HERA. It simultaneously incorporates constraints from global PDF fits and lattice-QCD inputs on GPDs at both zero and finite skewness. Overall, our results are systematically below the GUMP1.0 analysis, with an improving agreement observed at lower skewness.


\begin{figure}[htp]
\centering
\includegraphics[scale=.45]{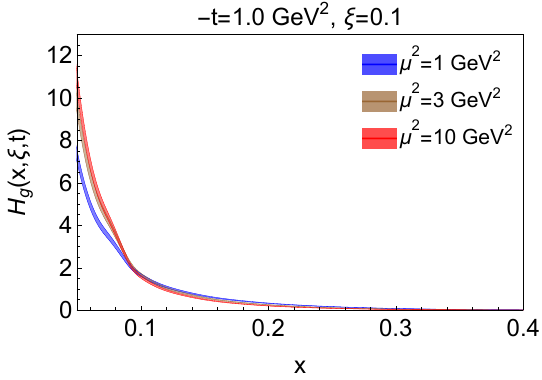}
\includegraphics[scale=.45]{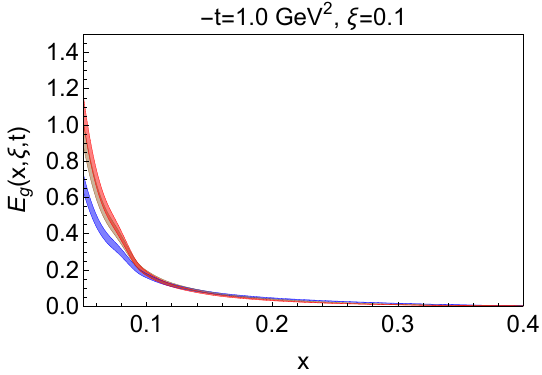}
\caption{Evolved gluon GPDs \(H(x,\xi,t)\) (left panel) and \(E(x,\xi,t)\) (right panel) at different scale $\mu^2=1,\,3,\,10$ GeV$^2$  shown as functions of \(x\) at fixed skewness \(\xi = 0.1\) and \(-t=1\) GeV$^2$.}
\label{gluon_full_evolved}
\end{figure}

Figure~\ref{gluon_full_evolved} displays the gluon GPDs at fixed momentum transfer $-t=1.0~\mathrm{GeV}^2$ and skewness $\xi=0.1$, evaluated at different scales. We observe that $H_g$ is significantly larger than $E_g$. This behavior arises because, at higher scales, the gluon GPDs are dominated by quark-to-gluon QCD evolution. For $H_g$, the contributions from $u$ and $d$ quarks have the same sign and therefore add constructively. In contrast, for $E_g$, the $u$- and $d$-quark contributions have opposite signs, with the $u$-quark contribution being larger in magnitude. As a result, these contributions partially cancel, leading to a much smaller $E_g$. Furthermore, both $H_g$ and $E_g$ increase in the small-$x$ region with increasing scale, as expected from QCD evolution.



\subsection{Compton form factors}
GPDs are directly connected to experimental observables such as cross sections and spin asymmetries, with DVCS providing a golden channel for their extraction. The DVCS amplitudes are parametrized in terms of CFFs, which are complex functions given by convolutions of hard coefficient functions with the corresponding GPDs.

At leading order and for factorization scale $\mu_F^2 = Q^2$, the CFFs associated with the chiral-even GPDs $H$ and $E$ are given by
\begin{equation}
\mathcal{F}(\xi,t)=\int_{-1}^{1} \mathrm{d}x
\left(
\frac{1}{x-\xi+i\epsilon}
+\frac{1}{x+\xi+i\epsilon}
\right)
F(x,\xi,t),
\end{equation}
where $F$ denotes either $H$ or $E$. The real and imaginary parts of $\mathcal{F}$ can be written explicitly as
\begin{equation}
\begin{aligned}
\mathrm{Re}\,\mathcal{F}(\xi,t) &= 
\mathcal{P}\!\int_{-1}^{1} \mathrm{d}x
\left(
\frac{1}{x-\xi}
+\frac{1}{x+\xi}
\right)
F(x,\xi,t),\\
\mathrm{Im}\,\mathcal{F}(\xi,t) &=
\pi\left[
F(\xi,\xi,t)-F(-\xi,\xi,t)
\right],
\end{aligned}
\end{equation}
where $\mathcal{P}$ denotes the Cauchy principal value.

\begin{figure}[htp]
\centering
\includegraphics[scale=.32]{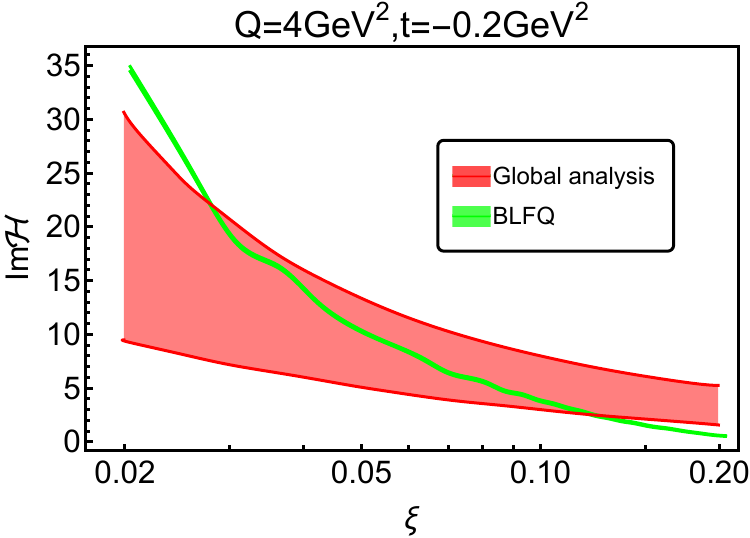}
\includegraphics[scale=.32]{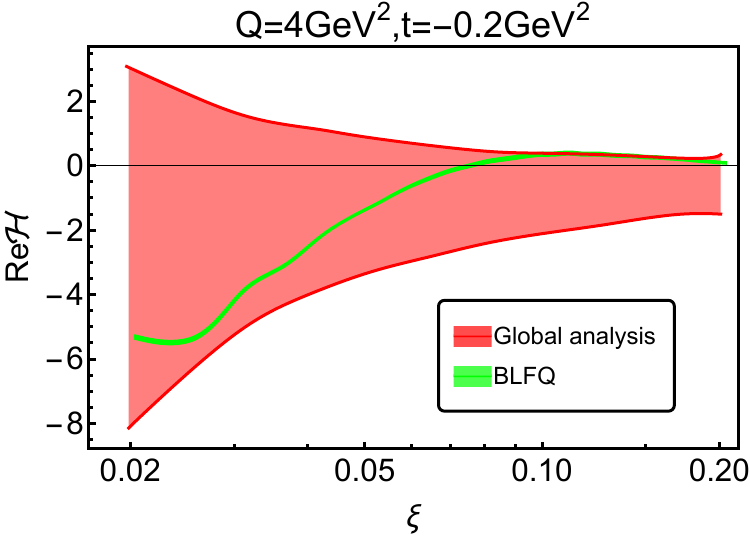}
\caption{Imaginary and real parts of the CFF $\mathcal{H}$ as function of $\xi$ at $Q^2 = 4~\mathrm{GeV}^2$ and $t = -0.2~\mathrm{GeV}^2$, compared with a neural-network–based global analysis~\cite{Moutarde:2019tqa}.}
\label{fig:CFFs}
\end{figure}

We use the evolved GPDs to compute the corresponding CFFs. 
Figure~\ref{fig:CFFs} compares our BLFQ results with a neural-network--based global analysis~\cite{Moutarde:2019tqa}. 
As shown in the left panel, our results are consistent with the experimental data in the range 
\(0.03 < \xi < 0.13\). 
For \(\xi < 0.03\), our predictions are larger than those from the global analysis, 
while for \(\xi > 0.13\) they become smaller. 
The right panel shows that the real part of the CFF is consistent with the experimental results.

\section{Summary}

To summarize, we have solved the light-front QCD Hamiltonian for the nucleon within a truncated basis that includes three Fock sectors: the three-quark, three-quark–gluon, and three-quark–quark–antiquark components, using the BLFQ approach. This allows us to obtain the nucleon LFWFs and to investigate its three-dimensional partonic structure.

We have computed the valence, gluon, and sea-quark GPDs at nonzero skewness and studied the quark GPDs in both the DGLAP and ERBL regions. The GPDs in our approach are defined at a low hadronic scale, where nonperturbative dynamics dominates. We subsequently apply QCD evolution to evolve the GPDs to higher momentum scales and compute the corresponding CFFs for comparison with global analyses.

We have found that the quark GPDs obtained within the BLFQ framework show only qualitative agreement at lower skewness with the first global extraction of GPDs based on experimental and lattice-QCD data at next-to-leading order accuracy, as provided by the GUMP1.0 analysis~\cite{Guo:2025muf}. Our results for the CFFs are also found to be consistent with the corresponding global analysis using neural network techniques~\cite{Moutarde:2019tqa}.

\section*{Acknowledgments}
We thank Jiatong Wu for useful discussions. This work is supported by the National Natural Science Foundation of China under Grant No. 12305095, No.12375143, and No. 12250410251, by the Gansu International Collaboration and Talents Recruitment Base of Particle Physics (2023-2027), by the Senior Scientist Program funded by Gansu Province, Grant No. 25RCKA008.
%
C. M. is supported by new faculty start up funding the Institute of Modern Physics, Chinese Academy of Sciences, Grants No. E129952YR0. 
X. Z. is supported by Key Research Program of Frontier Sciences, Chinese Academy of Sciences, Grant No. ZDBS-LY-7020, by the Natural Science Foundation of Gansu Province, China, Grant No. 20JR10RA067, by the Foundation for Key Talents of Gansu Province, by the Central Funds Guiding the Local Science and Technology Development of Gansu Province, Grant No. 22ZY1QA006, by international partnership program of the Chinese Academy of Sciences, Grant No. 016GJHZ2022103FN, by the Strategic Priority Research Program of the Chinese Academy of Sciences, Grant No. XDB34000000.
J. P. V. is supported by the Department of Energy under Grant No. DE-SC0023692.  A portion of the computational resources were also provided by Taiyuan Advanced Computing Center.

\biboptions{sort&compress}
\bibliographystyle{elsarticle-num}
\bibliography{ProtonGPDletterref.bib}
\end{document}